\documentclass{article}
\usepackage[T1]{fontenc} 
\usepackage[utf8]{inputenc} 
\usepackage{ismir,amsmath,cite,url}
\usepackage{graphicx}

\usepackage{color}

\usepackage{acronym}
\acrodef{dtw}[DTW]{Dynamic Time Warping}
\acrodef{hmm}[HMM]{Hidden Markov Model}
\acrodef{nmf}[NMF]{Non-Negative Matrix Factorization}
\acrodef{olda}[OLDA]{Ordinal Linear Discriminant Analysis}
\acrodef{msd}[MSD]{Music Structure Discovery}
\acrodef{msa}[MSA]{Music Structure Analysis}
\acrodef{ssm}[SSM]{Self-Similarity-Matrix}
\acrodef{mir}[MIR]{Music Information Retrieval}
\acrodef{ssl}[SSL]{Self-Supervised-Learning}
\acrodef{sa}[SA]{Self-Attention}
\acrodef{tl}[TL]{Triplet Loss}
\acrodef{fc}[FC]{Fully-Connected}
\acrodef{cqt}[CQT]{Constant-Q-Transform}
\acrodef{fc}[FC]{Fully-Connected}
\acrodef{bce}[BCE]{Binary-Cross-Entropy}
\acrodef{dnn}[DNN]{Deep Neural Network}
\acrodef{cnn}[ConvNet]{Convolutional Networks}
\acrodef{msd}[MSD]{Music Structure Discovery}
\acrodef{lms}[LMS]{Log-Mel-Spectrogram}
\acrodef{pcp}[PCP]{Pich-Class-Profile}
\acrodef{lsd}[LSD]{Laplacian Structural Decompositon}

\acrodef{hc}[HC]{hand-crafted}
\acrodef{dl}[DL]{deep learning}
\acrodef{af}[AF]{audio features}
\acrodef{ds}[DS]{detection system}

\acrodef{auc}[AUC]{Area Under the Curve ROC}

\renewcommand{\L}[0]{\mathcal{L}}
\newcommand{\X}[0]{\mathbf{X}}
\renewcommand{\S}[0]{\mathbf{S}_{ij}}
\newcommand{\hS}[0]{\mathbf{\hat{S}}_{ij}^{\theta}}
\newcommand{\e}[0]{\mathbf{e}^{\theta}}

\newcommand{\ddd}[2]{\frac{\partial #1}{\partial #2}}

\def\lbd{}	

\title{SSM-Net: feature learning for Music Structure Analysis using a Self-Similarity-Matrix based loss}

\twoauthors
  {Geoffroy Peeters} {LTCI, Télécom-Paris, IP-Paris \\ {\tt geoffroy.peeters@telecom-paris.fr}}
  {Florian Angulo} {LTCI, Télécom-Paris, IP-Paris \\ {\tt florian.angulo@telecom-paris.fr}}

\def\authorname{G. Peeters, F. Angulo}

\begin{document}

\maketitle

\begin{abstract}
In this paper, we propose a new paradigm to learn audio features  for \ac{msa}. 
We train a deep encoder to learn features such that the  \ac{ssm} resulting from those approximates a ground-truth \ac{ssm}.
This is done by minimizing a loss between both \ac{ssm}s.
Since this loss is differentiable w.r.t. its input features we can train the encoder in a straightforward way. 
We successfully demonstrate the use of this training paradigm using the \ac{auc} on the RWC-Pop dataset.
\end{abstract}

\section{Introduction}

\acf{msa} is the task aiming at identifying musical segments that compose a music track (a.k.a. segment boundary estimation) and possibly label them based on their similarity (a.k.a. segment labeling).
%
Over the years, systems for \ac{msa} have switched from
\begin{itemize}
	\itemindent=-15pt
	\itemsep=-3pt
	\item \acl{hc} \acl{ds} (checker-board-kernel~\cite{foote_automatic_2000} or DTW~\cite{Muller2013TASLPStructure}) applied to \acl{hc} \acl{af} (MFCC or Chroma)  
	\item to \acl{dl} \acl{ds} (boundary detection using ConvNet~\cite{ullrich_boundary_2014,grill_music_2015,cohen-hadria_music_2017}) applied to \acl{hc} \acl{af}, and recently 
	\item to  \acl{hc} \acl{ds} (checker-board-kernel) applied to deep learned  features~\cite{mccallum_unsupervised_2019,Wang2021ISMIRSupervised}.
\end{itemize}
Among the paradigms used to learn these features, metric learning using the triplet loss \cite{schroff_facenet_2015} has been the most popular, either using unsupervised learning \cite{mccallum_unsupervised_2019} or using supervised learning \cite{Wang2021ISMIRSupervised}.
In this paper, we propose a new paradigm to learn these features, which is more straightforward and less-computationally expensive (on a GPU Tesla P100-PCIE, training in about 1 hour  for our approach and 24 hours for \cite{mccallum_unsupervised_2019}).

\section{Proposal: SSM-Net}

\begin{figure*}[ht]
	\centerline{\includegraphics[trim=0cm 4cm 0cm 0cm,width=1.0\textwidth]{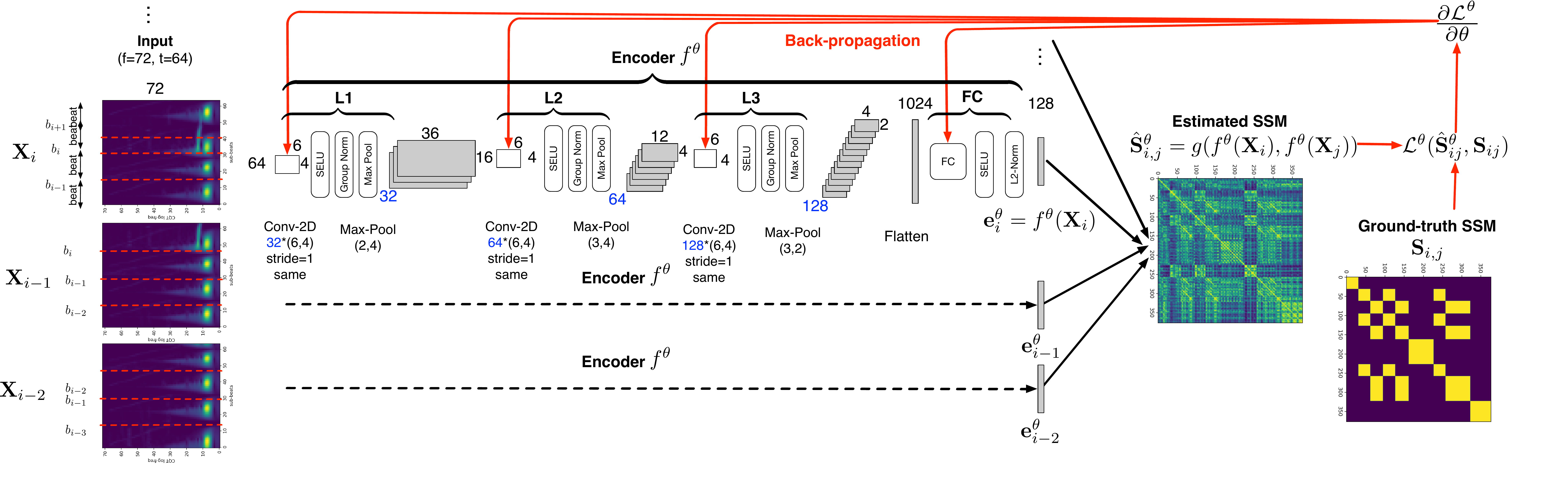}}
	\caption{SSM-net architecture. From left to right: input sequence $\{\X_i\}$ of beat-synchronous CQT-patches, encoder $f^{\theta}$ applied to each $\X_i$, estimated \ac{ssm} $\hS$ computed with embeddings $\{\e_i\}=f^{\theta}(\{\X_i\})$, Loss $\mathcal{L}^{\theta}$ computation.}
	\label{fig_smmnet}
\end{figure*}

Our SSM-net system is illustrated in Figure~\ref{fig_smmnet}.
The inputs and architecture (but not the loss) of our system are inspired by McCallum's  work~\cite{mccallum_unsupervised_2019} (but largely simplified\footnote{We reduced the sampling rate of the features by a factor 8: McCallum divides each beat into 128 sub-beats while we only use 16 sub-beats. We divided by a factor 2 the number of convolutional filters of each layer and we removed the last two fully connected layers.}).

\begin{figure*}[ht!]
	\centering
	\includegraphics[trim=6cm 2.5cm 0cm 0cm, width=1.05\textwidth]{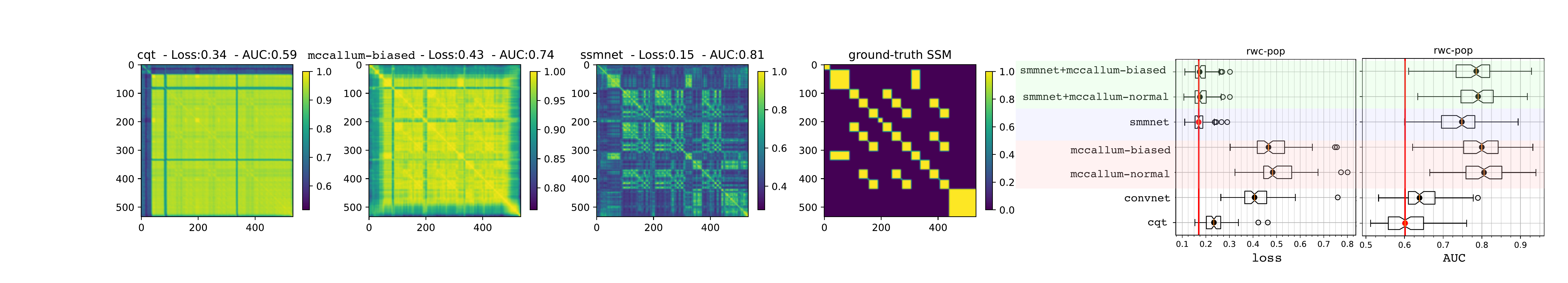}
	\caption{[Left] \ac{ssm}s $\hS$ computed using embeddings $\e$ obtained using (from left to right): \texttt{cqt}, \texttt{mccallum-biased}, \texttt{ssmnet} and ground-truth \ac{ssm} $\S$, on track 2 from RWC-Pop dataset. Loss $\mathcal{L}$ and AUC are indicated on top of each.
		[Right] Box-plots of Loss $\mathcal{L}$ and AUC obtained for all tracks of RWC-Pop dataset.}
	\label{fig_ssm-example}
\end{figure*}


\textbf{Input data $\{\X_i\}$.}
Each audio track is represented as a temporal sequence of $T$ audio features $\X_i$ which we denote as $\{\X_i\}_{i \in \{1 \ldots T\}}$ or $\{\X_i\}$ for short.
$\{\X_i\}$ are beat-synchronized patches of \ac{cqt}, each centered on a beat position $b_i$\footnote{The \ac{cqt}s are computed using \texttt{librosa}~\cite{mcfee2015librosa}. 
We used 72 log-frequencies ranging from C1 (31.70 Hz) to C7 (2093 Hz).}.
Each patch represents 4 successive beats\footnote{
The beat positions $\{b_i\}$ are computed using  \texttt{madmom} \cite{bock_enhanced_2011}\cite{madmom}.}. 
Each beat is further sub-divided into 16 sub-beats.
For this, the content of the \ac{cqt}s between two successive beats $b_{i-1}$ and $b_i$ is analyzed and clustered\footnote{using constrained agglomerative clustering and median aggregation as implemented in \texttt{librosa.segment.subsegment}.} into 16.
The inputs to our network are therefore patches $\X_i$ of \ac{cqt}, each of size (72 frequencies, 4*16 sub-beats) and centered on a beat $b_i$.


\textbf{Network architecture $\e_i = f^{\theta}(\X_i)$.}
The architecture of our encoder $f^{\theta}$  is illustrated in  Figure~\ref{fig_smmnet}.
It comprises 3 consecutive blocks (L1, L2, L3) of a 2D convolution followed by a SELU \cite{selu} activation, a 2D group normalization \cite{Wu2019GroupN} with 32 channels and a 2D max-pooling, 
The convolutional layers use a kernel size of (f=6, t=4)\footnote{f and t denotes the frequency and time dimensions} and the max-pooling layers use respectively kernel sizes of (2, 4), (3, 4) and (3, 2). 
The output is then passed to a single \ac{fc} layer of 128 units with a SELU activation.
The output is then L2-normalized and constitutes the embedding $\e_i=f^{\theta}(\X_i)$.
$\theta$ denotes the set of parameters to be trained (348.400 parameters).
For comparison the original McCallum~\cite{mccallum_unsupervised_2019 } network has 1.280.768.


\textbf{SSM-Net Loss.}
We apply the same encoder $f^{\theta}$ to each input $\X_i$. 
We then obtain the corresponding sequence of embeddings $\{\e_i\}_{i \in \{1 \ldots T\} }= f^{\theta}(\{\X_i\}_{i \in \{1 \ldots T\} })$.
We can then easily construct an estimated \ac{ssm}, $\hS$, using a distance/similarity $g$ function between all pairs of projections: 
\begin{equation}
	\hS = g(\e_i=f^{\theta}(\X_i), \e_j=f^{\theta}(\X_j)),\;\;\; \forall i, j
\end{equation}
$g$ is here a simple cosine-similarity which we scale to $[0,1]$:
\begin{equation}
	\hS=1 - \frac{1}{4} \lVert \e_i - \e_j \rVert_2^2 \;\;\; \in [0,1]
\end{equation}

It is then possible to compare $\hS$ to a ground-truth binary \ac{ssm}, $\S$.
We formulate this as a multi-class problem (a set of $T^2$ binary classifications)  and minimize the sum of \ac{bce} losses.
We compensate the class unbalancing  by using a weighting factor $\lambda$ computed as the percentage of 1 values in $\S$.
\begin{equation}
	\label{eq_wBCE}
	\L^{\theta}\!=\!-\!\sum_{i,j=1}^T\! (1\!-\lambda)\!\left[\S\!\log(\!\hS\!)\!\right]\!+\!\lambda\!\left[\!(\!1\!-\!\S\!) \!\log\!(\!1\!-\!\hS\!)\!\right]
\end{equation}

Since the computation of the \ac{ssm} $\hS$ is differentiable w.r.t. to the embeddings $\{\e_i \}$, we can compute $\ddd{\L^{\theta}}{\theta} $
\begin{equation}
	\ddd{\L^{\theta}}{\theta} 
	= \sum_{i,j=1}^T \ddd{\L^{\theta}}{\hS} \left( \ddd{\hS}{\e_i} \ddd{\e_i}{\theta} + \ddd{\hS}{\e_j} \ddd{\e_j}{\theta} \right)
\end{equation}


\textbf{Training.}
We minimize the loss using MADGRAD~\cite{defazio_adaptivity_2021} with a learning rate of $5\!\times\!10^{-4}$, a weight decay of $10^{-2}$ and  early-stopping.
The mini-batch-size $m$ (here defined as the number of full-tracks) is set to 6.

\textbf{Generating a ground-truth SSM $\S$.}
To generate $\S$, we rely on the homogeneity assumption, i.e. we suppose that all $t_i$ that fall within an annotated segment are identical since they share the same label.
If we denote by $\text{seg}(t_i)$ the segment $t_i$ belongs to and by $\text{label}(\text{seg}(t_i))$ its label, we assign the value   $\S=1$ if  $\text{label}(\text{seg}(t_i)) = \text{label}(\text{seg}(t_j))$.

\section{Evaluation}

To evaluate the quality of the features independently of the choice of a specific detection algorithm for \ac{msa}, we  directly compare the ground-truth $\S$ and the $\hS$ obtained using various choices for $\e$. 
For each choice, we measure the  obtained Loss $\mathcal{L}$ (lower is better) and \ac{auc} (higher is better) .
We conside the following features $\e$:
\begin{itemize}
	\itemindent=-15pt
	\itemsep=-3pt
	\item \texttt{cqt}: the flattened \ac{cqt} patches $\{ \X_i \}$
	\item \texttt{convnet:} the output of the un-trained (random weight) encoder $f^{\theta}$ applied to $\{ \X_i \}$
	\item \texttt{ssmnet}: the output of $f^{\theta}$ trained with SSM-Net
	\item \texttt{mccallum-normal/biased}: the output of  the same encoder $f^{\theta}$ but trained using the two unsupervised metric learning approaches described in \cite{mccallum_unsupervised_2019}
	\item \texttt{ssmnet-mccallum-normal/biased}: same as for \texttt{ssmnet} but $f^{\theta}$ is  pre-trained using \texttt{mccallum-normal/biased}
\end{itemize}

To train our SSM-Net, , we used  a sub-set of 695 tracks from the labeled dataset Harmonix~\cite{nieto_harmonix_2019}.
To train the unsupervised metric learning approach described in \cite{mccallum_unsupervised_2019}, we used a large unlabeled dataset from \texttt{YouTube} of 26.000 tracks from various genres.
The evaluation is performed on RWC-Pop \cite{goto_development_2004} labeled with AIST annotations \cite{Goto2006ISMIRAIST}).

In Figure~\ref{fig_ssm-example} [Left], we give an example of the \ac{ssm} $\hS$ obtained using the embeddings $\e$ learned by the most representative approaches.
On this example, \texttt{ssmnet} gives the $\hS$ with the highest contrast and the closest to the ground-truth. It gets a small $\mathcal{L}$=0.15 and a high AUC=0.81.

In Figure~\ref{fig_ssm-example} [Right], we represent the box-plots of $\mathcal{L}$ and AUC considering all tracks of RWC-Pop.
As one can see, the  SSM-net approach leads to the lowest $\mathcal{L}$.
However McCallum leads to a higher AUC than SSM-Net.
We therefore combine the SSM-Net training with a McCallum pre-training.
This then leads to both a low $\mathcal{L}$ and a high AUC .
 This is the approach we will develop in the future.

\clearpage

\end{document}